\documentstyle[prl,aps,twocolumn,psfig,floats]{revtex}

\begin{document}

\title{Negative heat capacities and first order phase transitions in nuclei}

\author{ L. G. Moretto, J. B. Elliott, L. Phair, and G. J. Wozniak }

\address{ Nuclear Science Division, Lawrence Berkeley National Laboratory,
Berkeley, CA 94720\\ }

\date{\today}

\maketitle

\begin{abstract}
Anomalous negative heat capacities have been claimed as indicators of
first order phase transitions in finite systems in general, and for
nuclear systems in particular.\ \ A thermodynamic approach allowing
for all $Q$ value terms is used to evaluate heat capacities in finite
van der Waals fluids and finite lattice systems in the coexistence
region.  Fictitious large effects and negative heat capacities are
observed in lattice systems when periodic boundary conditions are
introduced.\ \ Small anomalous effects are predicted for small drops
and for finite lattice systems.\ \ A straightforward application of
the analysis to nuclei shows that negative heat capacities cannot be
observed for $A>60$.
\end{abstract}


\narrowtext

The quest for discovery of the liquid to vapor phase transition in
nuclei has progressed along two lines.\ \ On one hand, the cluster
abundance has been studied as a function of mass and temperature and
the phase diagram has been generated \cite{elliott-02} in terms of
Fisher's theory of clusterization \cite{fisher-67}.\ \ On the other,
caloric curves have been analyzed to look for plateaus and breaks
possibly associated with the phase transition \cite{pochodzalla-95};
and fluctuations have been translated into heat capacities in the hope
of discovering negative values which are widely considered to be
indicators of phase coexistence in small systems
\cite{dagostino-00,dagostino-02}.

Regarding experimental caloric curves, it has been pointed out that
their interpretation hinges upon an unknown pressure-volume
relationship in the experiment, e.g. constant pressure or constant
volume \cite{moretto-96,elliott-00.3}.

Regarding the heat capacities, negative values are typically obtained
in lattice gas, Ising, or Potts model calculations for finite systems
\cite{hill-55,gross-97,chomaz-00}.\ \ The origin of these negative
values is generically attributed to the generation of surface with its
attendant energy cost, which is significant in small systems and which
disappears in the thermodynamic limit.\ \ However, with numerical
calculations, it is not clear how this effect actually comes about,
and if and how it may apply to actual systems like nuclei.\ \ The
issue is all the more interesting since claims of negative heat
capacities have been made for nuclear systems
\cite{dagostino-00,dagostino-02}.

In this paper we are going to investigate the subject of caloric
curves and heat capacities of finite systems in the coexistence
region and the underlying role of varying potential energies (``ground
states'') with system size on the basis of simple and very general
thermodynamical concepts, in the hope of obtaining solid and
unambiguous conclusions on these matters.

Our study applies to leptodermous (thin skinned) van der Waals-like
fluids and to models such as Ising, Potts, and lattice gas, which are
capable of reproducing their general features.\ \ We shall show that:
	\begin{enumerate}
	\item drops of leptodermous systems can indeed show negative
	      heat capacities \underline{as a slight effect} if
	      allowed to evaporate at constant pressure;
	\item finite lattice systems can also present negative heat
	      capacities \underline{as a slight effect} if open
	      boundary conditions are applied;
	\item finite lattices systems present negative heat capacities
	      \underline{as a much greater effect} if periodic
	      boundary conditions are applied, the main signal being
	      associated with the latter, very artificial conditions;
	\item nuclei cannot show negative heat capacities above $A
	      \approx 60$, while negative heat capacities are possible
	      for $A$ values below $60$.
	\end{enumerate}

\begin{figure} [ht]
        \centerline{\psfig{file=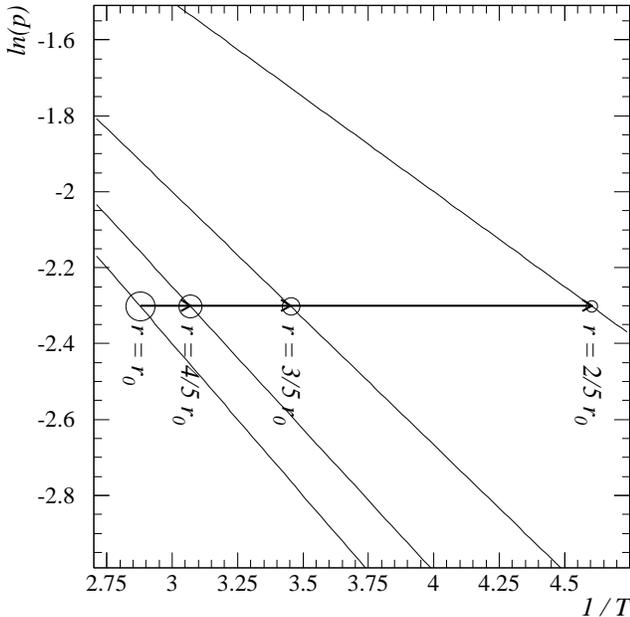,width=8.30cm,angle=0}}
        \caption{The natural log of the saturated vapor pressure as a
                 function of the inverse temperature for different
                 droplet radii.\ \ The size of the open circles is
                 proportional to the droplet radius.\ \ Arrows
                 illustrate the path of evaporation at constant
                 pressure.}
        \label{drop-coex}
        \end{figure}

Let us consider a macroscopic drop of a van der Waals fluid with $A$
constituents in equilibrium with its vapor.\ \ The vapor pressure $p$
at temperature $T$ is given by the Clapeyron equation
	\begin{equation}
	\frac{dp}{dT} = \frac{\Delta H_m}{T \Delta V_m} 
	\label{clap-eq}
	\end{equation}
where $\Delta H_m$ is the \underline{molar} vaporization enthalpy and
$\Delta V_m$ is the molar change in volume.\ \ The Clapeyron equation
gives a direct connection between what we might call the ``ground
state'' properties of the system and the saturation pressure along the
coexistence line.\ \ In fact, we can write
	\begin{equation}
	\Delta H_m = \Delta E_m + P \Delta V_m \sim \Delta E_m + T
	\label{ideal-enthalpy}
	\end{equation}
and for $T \ll \Delta E$, $\Delta H_m \approx a_v$ can be identified
approximately with the liquid-drop volume coefficient in the absence
of other terms like surface and Coulomb, etc.\ \ Assuming $\Delta V_m
\approx V_{m}^{\text{vapor}}$, that the vapor is an ideal gas and that
$\Delta H_m$ is constant with temperature, Eq.~(\ref{clap-eq}) can be
integrated to give
	\begin{equation}
	p \simeq p_o \exp \left(-\frac{\Delta H_m}{T} \right) .
	\label{pressure}
	\end{equation}
Equation~(\ref{pressure}) represents the $p$-$T$ univariant line in
the phase diagram of the system if $\Delta H_m$ is assigned its bulk
value $\Delta H^{0}_{m}$.\ \ It was observed long ago that for a drop
of finite size $\Delta H_m$ must be corrected for the surface energy
of the drop \cite{rayleigh}
	\begin{eqnarray}
	\Delta H_m & = & \Delta H_{m}^{0} + \Delta H_{m}^{s} 
		     = \Delta H_{m}^{0} - \gamma S_{m} \\ \nonumber
		   & = & \Delta H_{m}^{0} - a_{s} \frac{A^{2/3}}{A} 
	             = \Delta H_{m}^{0} - \frac{K}{r}
	\label{drop-enthalpy}
	\end{eqnarray}
where $\gamma$ is the surface tension, $S_m$ is the \underline{molar}
surface of the drop of radius $r$, $a_s$ is the surface energy
coefficient and $K$ is a geometrical constant.\ \ Substitution in
Eq.~(\ref{pressure}) leads to
	\begin{eqnarray}
	p   & = & p_0 \exp \left( -\frac{\Delta H_{m}^{0}}{T} 
              + \frac{a_s}{A^{1/3} T} \right) \\ \nonumber
	    & = & p_{\text{bulk}} \exp \left( \frac{a_s}{A^{1/3} T} \right)
	      = p_{\text{bulk}} \exp \left( \frac{K}{rT} \right) .
	\label{drop-pressure}
	\end{eqnarray}
At constant temperature the vapor pressure increases with decreasing
size of the drop.\ \ In other words, each drop of constant radius $r$
has its own $r$-dependent coexistence line with the vapor as shown in
Fig.~\ref{drop-coex}.

\begin{figure} [ht]
        \centerline{\psfig{file=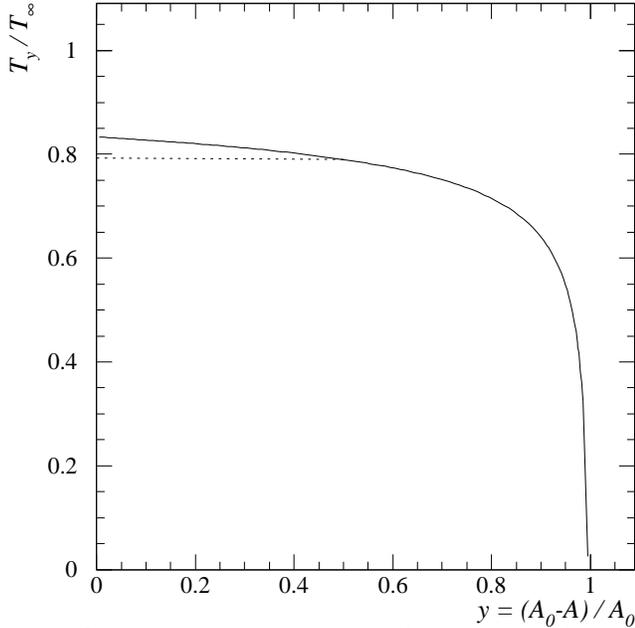,width=8.30cm,angle=0}}
        \caption{The temperature as a function of droplet size for a
                 drop evaporating at constant pressure in a system
                 with open boundary conditions.\ \ The solid line
                 shows the case of a spherical drop, while the dotted
                 line shows the case of a finite cubic lattice
                 evolving as in Fig.~\ref{drop-surface} top.}
        \label{drop-cp}
        \end{figure}

\begin{figure} [ht]
        \centerline{\psfig{file=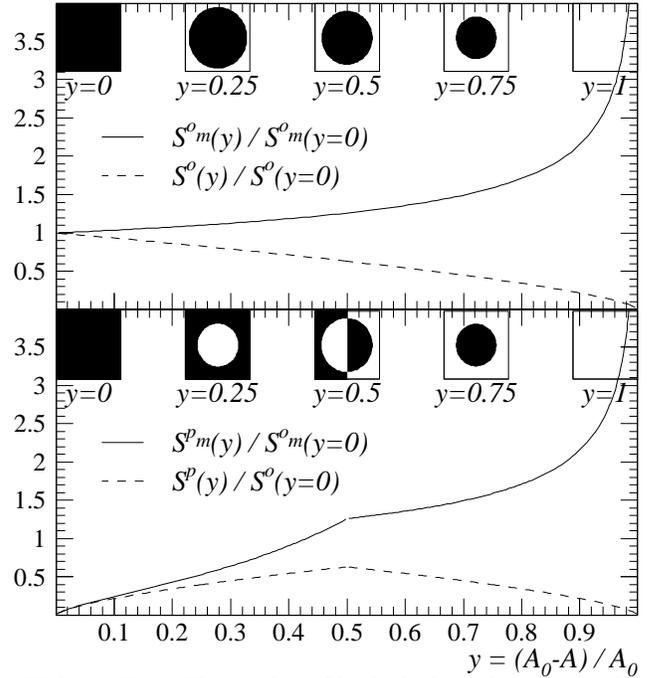,width=8.30cm,angle=0}}
        \caption{Top: The surface $S^o$ (dashed) and molar surface
                 $S^{o}_{m}$ (solid) area of a drop for open boundary
                 conditions normalized to their values at $y=0$.\ \
                 Bottom: The surface $S^p$ (dashed) and molar surface
                 $S^{p}_{m}$ (solid) area of a drop for periodic
                 boundary conditions normalized to their values at
                 $y=0$.\ \ In-sets show the configurations at various
                 values of $y$.}
        \label{drop-surface}
        \end{figure}

Let us now consider the case of isobaric evaporation of a drop
starting from a drop with $A_0$ constituents and evaporating into a
drop with $A <A_0$ constituents.\ \ Let us now define the drop size
parameter
	\begin{equation}
	y = \frac{A_0 - A}{A_0} .
	\label{drop-size}
	\end{equation}
At constant pressure
	\begin{equation}
	p_0 \exp \left( -\frac{\Delta H_{m}^{0}}{T} \right) = 
	p_0 \exp \left( -\frac{\Delta H_{m}(y)}{T_y} \right).
	\label{drop-const-pres}
	\end{equation}
Remembering that in most liquids 
$\left| a_v \right| \approx \left| a_s \right|$ and that in the lattice gas 
$\left| a_v \right| =       \left| a_s \right|$, we have
	\begin{equation}
	\Delta H_m (y) \simeq a_v \left( 1 - \frac{1}{A_{0}^{1/3}(1-y)^{1/3}}\right)
	\label{drop-size-enthalpy}
	\end{equation}
from which follows
	\begin{equation}
	\frac{T_y}{T_{\infty}} \simeq \frac{\Delta H_m (y)}{\Delta H_{m}^{0}} 
	              \simeq 1 - \frac{1}{A^{1/3}}
	              \simeq 1 - \frac{1}{A_{0}^{1/3}(1-y)^{1/3}} .
	\label{red-temp}
	\end{equation}
Thus, a slight \underline{decrease} in temperature is predicted as the
drop evaporates isobarically, thus leading to a negative isobaric heat
capacity in the coexistence region as illustrated in
Fig.~\ref{drop-cp}.\ \ In this figure, the abscissa is trivially
related to $\Delta H$, the heat absorbed by the evaporating drop at
constant pressure, by the relationship
	\begin{equation}
	\Delta H = A_0 \int_{0}^{y} dy \Delta H_m (y) = A_0 y \overline{\Delta H_m} .
	\label{trival}
	\end{equation}
The same decrease can be more visually appreciated from
Fig.~(\ref{drop-coex}).\ \ As the drop is evaporating at constant
pressure, the drop moves from one coexistence curve to another
according to its decrease in radius, and thus to progressively lower
temperatures.\ \ This slight effect is due \underline{not} to an
increase in surface as the drop evaporates, since the drop surface of
course \underline{diminishes} as $A^{2/3}$, but to the slight increase
of molar surface which \underline{does} increase as $A^{-1/3}$ as
shown in Fig.~\ref{drop-surface}.\ \ Also, the formation of bubbles in
the body of the drop is thermodynamically disfavored by the factor $f
= \exp (-\gamma \Delta S / T)$ where $\Delta S$ is the surface of the
bubble.

It is worth pointing out that the smallness of the effect ($\sim 4$\%
decrease in $T$ as a droplet of $A \sim 200$ evaporates all the way to
$A \sim 100$), is needlessly magnified by the widely practiced artful
translation of the caloric curve into heat capacity, which jumps from
$+\infty$ to $-\infty$ as the slope of the caloric curve changes from
constant to a slightly negative value.

It might be argued that cluster formation may be responsible for
negative heat capacities.\ \ In physical fluids as well as in the
Ising model the ``mean'' molecular weight of the equilibrium vapor
remains very close to that of the monomer, as confirmed by the
adherence of the vapor pressure vs. temperature to the
Clapeyron-Clausius formula with constant $\Delta H_m$
\cite{gugg-text}.\ \ Furthermore, if such an effect existed it would
operate already for the infinite systems, which, patently, it does not.


Let us now move to the amply studied cases of lattice gas, Ising, and
Potts models.\ \ Here as above, the study of the evolution of the
ground state properties can be directly translated into the
liquid-vapor coexistence properties.\ \ We consider first an
evaporating finite system in three dimensions of size $A_0 = L^3$,
with open boundary conditions.\ \ This case is essentially identical
to the case of a drop discussed above (see Fig.~\ref{drop-cp}).

For maximal density at $T=0$ (the ground state) $y=0$ and the entire
cubic lattice is filled.\ \ For decreasing densities, always at $T=0$
a single cluster of minimum surface is present, which evolves from a
cube to a sphere.\ \ The associated change in surface is shown in
Fig.~(\ref{drop-surface}).

Since a cube has only a slightly larger surface than a sphere of the same
volume by a factor of $6 / \left( \left( 4 \pi \right) ^{1/3} 3^{2/3}
\right)$ the resulting decrease of $T_y$ at $y=1$ is nearly exactly
what a sphere of the same volume would experience in going from $y=0$
to $y = 1/2$ ($A_0 \sim 200$).\ \ Thus the caloric curve from $y=0$ to
$y=1/2$ is essentially \underline{flat} like in the infinite system,
and the heat capacity is trivially infinite.

So, where are the large effects reported in so many papers
\cite{hill-55,gross-97,chomaz-00}?\ \ We shall see below the effect
created by the introduction of periodic boundary conditions.


The introduction of periodic boundary conditions rids the system of
``dangling bonds,'' as it were, by repeating a cubic lattice of side
$L$ periodically along the three coordinates.\ \ These conditions,
originally introduced to mimic the infinite system, lead here to
peculiar consequences.

At $y=0$, the lattice is filled with particles so that $\Delta H_m(0)
= \Delta H_{m}^{0}$ characteristic of the infinite system.\ \ As $y$
increases at fixed lattice size, a \underline{bubble} develops in the
cube and \underline{surface} is rapidly created.\ \ This is shown in
Fig.~(\ref{drop-surface}).\ \ The bubble develops since the periodic
boundary conditions prevent 'as it were' evaporation from the
surface.\ \ The bubble grows with increasing $y$ until it touches the
sides of the lattice.\ \ This occurs for $y \approx 1/2$.\ \ At nearly
$y=1/2$ and beyond, the ``stable'' configuration is a drop that
eventually vanishes at $y=1$.\ \ The change in surface associated with
the range $0 \le y \le 1$ as well as the molar surface are shown in
the bottom panel of Fig.~\ref{drop-surface}.

        \begin{figure} [ht]
        \centerline{\psfig{file=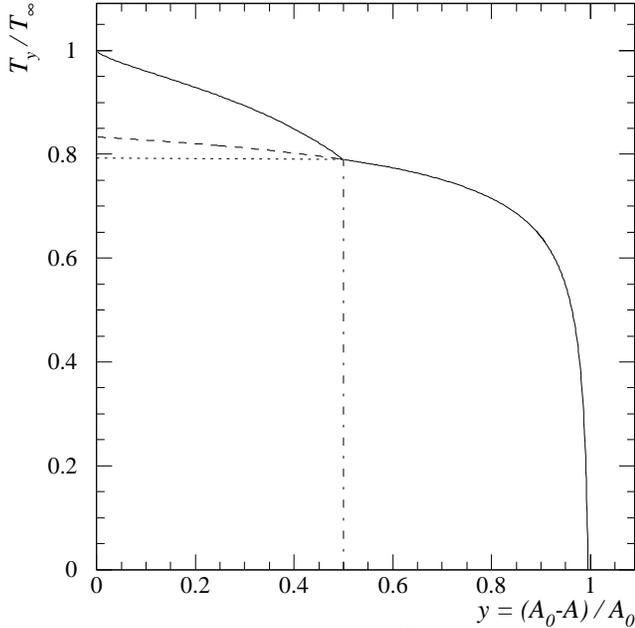,width=8.30cm,angle=0}}
        \caption{The temperature as a function of droplet size for a
                 drop evaporating at constant pressure in a system
                 with periodic boundary conditions.\ \ The solid line
                 shows the case of a finite cubic lattice with
                 periodic boundary conditions evolving as in
                 Fig.~\ref{drop-surface} bottom, while the dotted line
                 and the dashed line are the same as in
                 Fig.~\ref{drop-cp} and the vertical dash-dotted line
                 indicates the case of $50$\% lattice occupation.}
        \label{drop-lg-pbc}
        \end{figure}

The evaporation enthalpy thus becomes
	\begin{equation}
	\Delta H_m(y) \simeq 
	a_v 
	\left( 
		1 - \frac{y^{2/3}} {A_{0}^{1/3}\left( 1 - y \right)} 
	\right)
	\label{lg-pbc-enthalpy1}
	\end{equation}
from $y=0$ to $y=1/2$, and
	\begin{equation}
	\Delta H_m(y) \simeq a_v \left( 1 - 
	\frac{1}{A_{0}^{1/3}\left( 1 - y \right)^{1/3}} \right)
	\label{lg-pbc-enthalpy2}
	\end{equation}
from $y=1/2$ to $y=1$.

As a consequence, for periodic boundary conditions
	\begin{equation}
	\frac{T_y}{T_{\infty}} \simeq 1 
	- \frac{y^{2/3}}{A_{0}^{1/3}\left( 1 - y \right)}
	\label{lg-pbc-temp}
	\end{equation}
from $y=0$ to $y=1/2$, while from $y=1/2$ to $y=1$
Eq.~(\ref{red-temp}) holds.

The results contained in equations (\ref{red-temp}) and
(\ref{lg-pbc-temp}) are very general and convenient.\ \ They bypass
the clumsy numerical calculations for individual systems, and require,
for any case, just the knowledge of the geometry and of the surface
energy coefficient.

The dramatic effect of periodic boundary conditions can now be seen in
Fig.~\ref{drop-lg-pbc}.\ \ The temperature decreases substantially
with increasing $y$, due to the fact that the molar enthalpy at $y=0$
assumes its bulk value $\Delta H_{m}^{0}$ and must meet the previous
case of open boundary conditions for $y=1/2$.\ \ This may well explain
the calculated negative heat capacities reported in literature, as an
artifact due to the unnatural choice of boundary conditions.\ \ The
conclusion of this exercise is the following: if $\Delta H_m (y)$
decreases with $y$ (with decreasing drop size), for any reason, we
expect negative heat capacities.\ \ Alternatively we expect the normal
infinite heat capacities if $\Delta H_m =$ constant or positive heat
capacities if $\Delta H_m$ increases with $y$.

\begin{figure}[ht]
        \centerline{\psfig{file=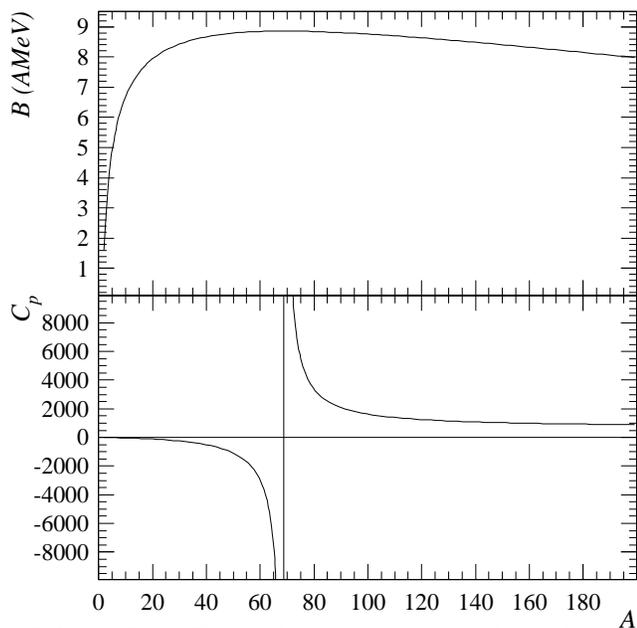,width=8.30cm,angle=0}}
        \caption{Top: The binding energy per nucleon of atomic
                 nuclei.\ \ Bottom: The associated heat capacity.}
        \label{ebind}
        \end{figure}

With the lessons learned above we can evaluate the heat capacities for
nuclei.\ \ It is apparent from the above arguments that the key
quantity is $\Delta H_m$ and its dependence on the drop size,
irrespective of the physical causes that determine its magnitude and
dependence.\ \ In the case of nuclei the quantity $\Delta H_m$ is
determined not only by the volume and surface, but also by all the
other terms in the liquid drop model, such as the Coulomb and symmetry
energy all of which contribute to the mean binding energy per nucleon.\
\ Consequently one can immediately infer that when the binding energy
per nucleon \underline{decreases} with $A$, the heat capacity should
be positive, and vice-versa.\ \ Thus, since the maximum binding energy
per nucleon occurs at $A \sim 60$, negative heat capacities should be
possible only for $A < 60$.\ \ Let us proceed more precisely.\ \ We
can rely again on the Clapeyron equations to calculate the heat
capacity as follows
	\begin{equation}
	C_p =   \left. \frac{dH}{dT} \right|_p
	    = - \left. \frac{dH}{dA} \right|_p 
                \left. \frac{dA}{dT} \right|_p
	    = - \left. \Delta H_m (A) \frac{dA}{dT} \right|_p
	\label{nuclei-cp1}
	\end{equation}
but
	\begin{equation}
	\left. \frac{dT}{dA} \right|_p = 
	\left. \frac{dp}{dA} \right|_T 
	\left. \frac{dT}{dp} \right|_p .
	\label{step1}
	\end{equation}
From the integrated form of the Clapeyron equation we have
	\begin{equation}
	\left. \frac{dp}{dA} \right|_p = - \frac{1}{T}\frac{d \Delta H_m}{dA}p
	\label{step2}
	\end{equation}
so
	\begin{equation}
	\left. \frac{dT}{dA} \right|_T = 
	- \frac{1}{T}\frac{d \Delta H_m}{dA}p
	\frac{T V_m}{\Delta H_m} =
	-\frac{T}{\Delta H_m} \frac{d \Delta H_m}{dA} .
	\label{step4}
	\end{equation}
Finally
	\begin{equation}
	C_p = \frac{\frac{\left( \Delta H_m (A) \right)^2}{T}}
	    {\frac{d \Delta H_m}{dA}} .
	\label{nucl-cp2}
	\end{equation}
The derivative in the denominator can be evaluated approximately from
the dependence on the binding energy per nucleon $B$ upon the mass
number
	\begin{equation}
	\frac{d \Delta H_m}{dA} = \frac{dB}{dA} .
	\label{enthalpy-Ebinding}
	\end{equation}
The liquid drop model allows us to estimate such a derivative.\ \
Without the Coulomb term, of course, we recover the results presented
above for a drop: the binding energy increases with increasing $A$ and
tends asymptotically to the value $a_v \approx 15$MeV.\ \ Thus
negative heat capacities should be expected and the caloric curves
should look like that shown in Fig.~\ref{drop-cp}.\ \ The Coulomb and
symmetry terms, however, become very important at large values of $A$,
say, along the line of $\beta$-stability.\ \ From Fig.~\ref{ebind} it
is apparent that the binding energy \underline{decreases} with $A$ for
$A>\sim60$.\ \ Consequently in all this region of $A$, positive
specific heats should be expected.\ \ Only for $A<\sim60$, negative
specific heats are predicted.

Despite the assumption of a temperature independent $\Delta H_m$ we
expect these results to hold over a very broad range of temperatures.\
\ On one hand a monotonic change of $\Delta H_m$ with $T$ should not
alter the result; on the other hand it is known that for van der Waals
fluids their vapor pressure $p = p(T)$ is well defined up to the
critical point by a constant $\Delta H_m$ \cite{gugg-text}.\ \
Furthermore the appearance of clusters has minimal influence on
$\Delta H_m$ and therefore on our results, because the vapor
concentration is dominated by monomers \cite{elliott-02}.

This straightforward result based on elementary thermodynamics and
ground state binding energies raises serious questions as to the
meaning of the negative heat capacities that have been claimed for
large nuclear systems.

In conclusion, we have shown that:
	\begin{enumerate}
	\item it is possible to generate caloric curves and heat
	      capacities in the coexistence region from the knowledge
	      of the molar heat of vaporization, which must include
	      all $Q$-value terms;
	\item simple drops and Ising, Potts models with open boundary
	      conditions show minimal anomalies, while periodic
	      boundary conditions introduce strong anomalies as
	      artifacts;
	\item nuclei with $A > \sim 60$ should present no anomalous
	      negative heat capacities while this is possible for $A
	      \le \sim 60$.
	\end{enumerate}

This work was supported by the Director, Office of Energy Research,
Office of High Energy and Nuclear Physics, Nuclear Physics Division of
the US Department of Energy, under Contract No. DE-AC03-76SF00098.

\end{document}